\documentclass[letterpaper]{article}    

\begin{document}

\bibliographystyle{unsrt}    

\newtheorem{theorem}{Theorem}
\newtheorem{definition}{Definition}
\newtheorem{cor}{Corollary}
\newtheorem{lem}{Lemma}

\title{Generalized Stable Multivariate Distribution and Anisotropic Dilations}

\author{D. Schertzer
,
M. Larchev\^eque \\
Laboratoire de Mod\'elisation en M\'ecanique, Tour 66, Boite 162
\\ Universit\'e
Pierre et Marie Curie,\\ 4 Place Jussieu F-75252 Paris Cedex 05, France.\\
\and
J. Duan \\ Department of Mathematical Sciences \\
Clemson University \\
Box 341907, Clemson, SC 29634-1907, USA.\\
\and
S. Lovejoy \\ Physics Department, McGill U.\\
3600 University Street
Montreal, H3A 2T8, Quebec, Canada.}
\date{submitted to J. Multivariate Anal., July 27 1999}

\maketitle

\abstract{After having  closely re-examined the notion of a L\'evy's stable
vector,
it is shown that the notion of a stable multivariate distribution is more
general
than previously defined. Indeed, a more intrinsic vector definition is
obtained with the help of non isotropic
dilations and a related notion of generalized scale.
In this framework, the components of a stable vector
may not only have distinct Levy's stability indices $\alpha$'s, but the
latter may depend on its norm.
Indeed, we demonstrate that the Levy's stability index of a vector rather
correspond to a linear application
than to a scalar, and we show that the former should satisfy a simple
spectral property.}

\section {Introduction and notation}

In order to define Levy stable (or $\alpha$-stable)
vectors valued on a real Hilbert space $H$
and their corresponding multivariate distribution in a more general
way than the classical and nearly isotropic definition \cite{Levy},
we will use a notion of scale $\parallel \underline{x} \parallel$ which, as
discussed in the following section, is weaker than the
canonical norm $\mid\underline{x}\mid$
defined by the scalar product  $(\underline{x}, \underline{y})$;

\begin{equation}\label{eq.norm}
\mid\underline{x}\mid ={( \underline{x}, \underline{x})}^{1/2}
\end{equation}

Let $L(H,H)$ denote the set of (continuous) endomorphisms of $H$. Although
most of the properties discussed below do not depend on the precise
structure of  $H$, it
can be considered as being $H= R^d$.
For the Laplace-Fourier transform, we need to consider its complexified
space $\overline{H}$
($\overline{H}= H +iH$) equipped with the hermitian extension of the scalar
product:

\begin{equation}
 \forall \underline{q} \in \overline{H}, \forall \underline{x} \in H: \quad
(\underline{q},\underline{x}) = (Re(\underline{q}),\underline{x})
+i(Im(\underline{q}),\underline{x})
\end{equation}

\noindent whereas for a Fourier transform it suffices to consider $iH$, and
for a Laplace transform
a subspace $H^+ \subset H$, e.g. $H^+=(R^+)^d$ for $H=R^d$.

For any $\underline{\underline{M}} \in  L(H,H)$,
$Spec(\underline{\underline{M}})$ denotes its spectrum,
$Re(Spec(\underline{\underline{M}})$
its real part, i.e. the set of the real parts of its eigenvalues,
$Ker(\underline{\underline{M}})$
its kernel,
$sym(\underline{\underline{M}})$ its symmetric part, i.e.
$2.sym(\underline{\underline{M}})= \underline{\underline{M}} +
\underline{\underline{M}}^*$, where
$\underline{\underline{M}}^*$ denotes the hermitian conjugate of
$\underline{\underline{M}}$.

\section {A generalized notion of scale}\label{sect_GSI}

The need for a generalized notion of scale will be more apparent in
sect.\ref{sect_characteristic},
however its definition is rather straightforward \cite{gsi}, at least for
the linear case which will be
sufficient for the following. However, the subsequent developments can be
skipped in a first reading.

\begin{definition}\label{def_GSI}
An application $\parallel.\parallel$ from a real Hilbert space $H$ onto the
set of positive real numbers $ R^+$
is said to be  a  generalized notion of scale, associated to one-parameter
(linear)
dilation $\underline{\underline{T}}_{\lambda}$, when it satisfies the
following:
\end{definition}
\begin{itemize}

\item{\it nondegeneracy}, i.e:
\end{itemize}
\begin{equation}\label{eq.nondegeneracy}
\parallel \underline{x} \parallel=0 <=> \underline{x} = \underline{0}
\end {equation}

\begin{itemize}
\item{\it linearity with the dilation parameter}, i.e:

\end{itemize}

\begin{equation}\label{eq.linearity}
\forall \underline{x} \in H, \forall \lambda \in R^+ :\quad
\parallel {\underline{\underline{T}}_{\lambda}.\underline{x}  }\parallel=
\lambda \parallel\underline{x}\parallel
\end{equation}

\begin{itemize}
\item{\it  Balls defined by this scale are strictly increasing}, i.e.:
\end{itemize}

\begin{equation}\label{linearity}
\forall \lambda, \lambda'\in R^+; \lambda \geq \lambda': \quad
 {B_{\lambda}} \supseteq {B_{\lambda'}}
\end{equation}

\noindent where the balls $B_{\lambda}$ defined by the dilation
$\underline{\underline{T}}_{\lambda}$ satisfy (due to eq.\ref{eq.linearity}):

\begin{equation}\label{balls}
B_{\lambda} \equiv \underline{\underline{T}}_{\lambda}(B_{\lambda}) =
{\{ \underline{x}\in H; \parallel\underline{x}\parallel \leq \lambda \} }
\end{equation}

\noindent and have the following frontier:
\begin{equation}\label{frontier_balls}
\partial B_{\lambda}= \{ \underline{x}\in H;
\parallel\underline{x}\parallel = \lambda \}
\end{equation}

It is straightforward to check that the canonical norm $\mid.\mid$
is the scale associated to
the isotropic dilation $\underline{\underline{T}}_{\lambda}=
 \lambda \underline{\underline{1}}$.
Whereas the two first properties are rather identical to
those of a norm,
the last one is weaker than the triangular inequality which is required for
a norm.

The conditions of existence of a generalized scale should depend on the
generator $\underline{\underline{G}}$
of the one parameter group of (linear) dilations
$\underline{\underline{T}}_{\lambda}$:

\begin{equation}\label{eq.generator}
\underline{\underline{T}}_{\lambda}=
{\lambda}^{\underline{\underline{G}}}
= e^{Ln \lambda \underline{\underline{G}}}
\end{equation}

Indeed, we have the following theorem \cite{gsi}:

\begin{theorem}\label{theo_existence_gsi}
Let us consider the unit ball defined as an ellipsoid generated by a
positive symmetric matrix
$\underline{\underline{A}}$:

\begin{equation}\label{eq.symm.A}
B_1 = {\{ \underline{x}\in H; (\underline{x},
\underline{\underline{A}}.\underline{x})^{1/2} \leq 1 \} }
\end{equation}

The dilation group $\underline{\underline{T}}_{\lambda}$ defines a
generalized scale, if and only if
its generator $\underline{\underline{G}}$ satisfies:

\begin{equation} \label{eq_theo_existence_gsi}
Spec(sym(\underline{\underline{A}}.\underline{\underline{G}}) )\geq 0
\end{equation}

\end{theorem}

Indeed, the balls $B_{\lambda}$ are then defined by:

\begin{equation}
\underline{\underline{A}}_{\lambda} =
(\underline{\underline{T}}^{-1}_{\lambda})^*.
\underline{\underline{A}}.\underline{\underline{T}}^{-1}_{\lambda} :~~
B_{\lambda} = {\{ \underline{x}\in H; (\underline{x},
\underline{\underline{A}}_{\lambda}.\underline{x})^{1/2} \leq 1 \} }
\end{equation}

\noindent and we have:
\begin{equation}
{d \over {d\lambda}} (\underline{x},
\underline{\underline{A}}_{\lambda}.\underline{x}) =
- (\underline{\underline{T}}_{\lambda}^{-1}.\underline{x},
sym(\underline{\underline{A}}.\underline{\underline{G}}).
\underline{\underline{T}}_{\lambda}^{-1}.\underline{x})
\end{equation}

This theorem has the following corrolary, which will be sufficient for the
following:

\begin{cor}\label{cor_existence_gsi}
When the unit ball is an ellipsoid defined by a positive symmetric matrix
$\underline{\underline{A}}$ (eq.\ref{eq.symm.A}),
the dilation group $\underline{\underline{T}}_{\lambda}$ defines a
generalized scale,
if and only if its generator $\underline{\underline{G}}$ satisfies:

\begin{equation} \label{eq_cor_existence_gsi}
Spec(sym(\underline{\underline{G}})) \geq 0 \Longleftrightarrow
Re(Spe(\underline{\underline{G}})) \geq 0
\end{equation}

\noindent and when $\underline{\underline{A}}$ belongs to a given
neighbourhood of a scalar linear application, i.e.
$\exists \mu \in R^+: \underline{\underline{A}} = \mu
\underline{\underline{1}}$.
\end{cor}

Indeed, for the scalar case ($\underline{\underline{A}} = \mu
\underline{\underline{1}}$), it is
a straightforward consequence of theo.\ref{theo_existence_gsi}. Due to
continuity, it is also valid
for a given neighbourhood of this case, i.e. when the eigenvalues of
$\underline{\underline{A}}$
are not too much different. However, simple counter examples (e.g. in
$L(R^2,R^2)$) are easy to find, i.e.
generators  $\underline{\underline{G}}$ which satisfy
eq.\ref{eq_cor_existence_gsi}, but violate
eq.\ref{cor_existence_gsi} as soon as the eigenvalues of
$\underline{\underline{A}}$
are different enough \cite{Gao}. This shows that this neighbourhood is
indeed bounded.

\section {Levy stable vectors}

Let us recall that a (real) Levy stable variable \cite{Levy, Feller,
Zolotarev, Samarodinsky} can be defined
in the following manner, e.g. \cite{Feller} ($=^d$ denotes the equality in
distribution):

\begin{definition}\label{def1_levy_variable}
A random variable $X$ is said to be a Levy stable variable, iff it is
stable under renormalized sum
--i.e. with the rescaling factor $a(n)$ and recentring term $\gamma(n)$--
of any  $n$ of its independent realizations $ X_i, (i=1,n)$. This
corresponds to:

\begin{equation}\label{eq1_levy_variable}
 X_i =^d X (i=1,n):~ \\
\forall n \in N, \exists a(n),\gamma(n) \in R \\
\sum_{i=1}^n X_i =^d  a(n) X + \gamma(n)
\end{equation}
Furthermore, $X$ is said {\em strictly} stable \footnote{according to
Feller's terminology}
when the recentring term $\gamma(n)$ is $0$.
\end{definition}

It is straightforward to check (by induction) that this definition is
equivalent
to the original definition given by Levy \cite{Levy}
which addresses the stability under any linear combination:

\begin{definition}\label{def2_levy_variable}
Two identically distributed random variables $X_1, X_2$ are said to be stable
under linear combination iff:

\begin{equation}\label{eq2.levy_variable}
X_1 =^d X_2=^d X ~:~~ \\
\forall a_1,a_2\in R^+, \exists a \in R^+, \gamma \in R:  \\
a_1 X_1 + a_2 X_2 =^d  a X + \gamma \\
\end{equation}

Furthermore, they are said {\em strictly} stable, when the recentring term
$\gamma $ is $0$.

\end{definition}

Due to their linearity, these definitions can be extended in a
rather straightforward manner to random vectors \cite{Levy, Paulauskas,
Nikias}, with the only necessary
modification that the (rather trivial) recentring term
($\gamma(n)$ in eq.\ref{eq1_levy_variable}, $\gamma$ in
eq.\ref{eq2.levy_variable})
is now a vector \footnote{In fact, this definition might have seemed so
trivial to  L\'evy
that it is not explicitely written in \cite{Levy}}.
However, this extension seems rather
restrictive, one of its consequences, recalled below, is that all the
components have
the same Levy index. We will  therefore consider the following more
intrinsic vector definition of a stable vector:

\begin{definition}\label{def_levy_vector}
A random vector $\underline{X}$  valued on an Hilbert $H$ is said to be a
Levy stable vector,
iff it is stable under renormalized sum --i.e. with a rescaling linear
application
$\underline{\underline{a}}(n) $ and recentring vector $\underline{\gamma}(n)$--
of any  $n$ of its independent realizations $ \underline{X}_i, i=1,n$. This
corresponds to:

\begin{equation}\label{eq_levy_vector}
 \underline{X}_i =^d  \underline{X} (i=1,n):~~ \\
\forall n \in N, \exists  \underline{\underline{a}}(n) \in L(H,H),
\underline{\gamma}(n) \in H: \\
\sum_{i=1}^N \underline{X}_i =^d  \underline{\underline{a}}(n) .
\underline{X} +
\underline{\gamma}(n)
\end{equation}
Furthermore,  $\underline{X}$ is said {\it strictly} stable,
when the recentring term $\underline{\gamma}(n)$ is $\underline{0}$.
\end{definition}

The classical definition \cite{Levy, Paulauskas, Nikias} corresponds to the
scalar case for \underline{\underline{a}}(n):

\begin{equation}\label{eq_classical_levy_vector}
\underline{\underline{a}}(n) = a(n) \underline{\underline{1}}
\end{equation}

\noindent and therefore can be called 'quasi-scalar case' of L\'evy stable
vectors.

\section {Attractivity}

We can now extend the definition of attractivity to random  vectors :

\begin{definition}\label{def_attractive_vector}
The distribution $R$ of the independent random vectors $ \underline{X}_i$
belongs to the
domain of attraction of a distribution $R_a$ of the random vector $
\underline{X}$,
if the distribution of their renormalized sum --i.e.with a rescaling linear
application
$\underline{\underline{a}}(n) $ and recentring vector $\underline{b}(n)$--
 tends to $R_a$. This corresponds to:
\end {definition}

\begin{equation} \label{eq_attractive_vector}
lim_{n \to \infty} { {\sum_{i=1,n} \underline{X}_i -\underline{b}(n)} \over
{\underline{\underline{a}}(n)}}
=^d   \underline{X}
\end{equation}

By its very definition (def.\ref{def_levy_vector}) each stable vector
belongs to
its own domain of attraction and we will show there is no other possible
limit.
This will demonstrate that the stable vectors are the only attractive
vectors.

\section {Scaling law of the rescaling factor}

As for the scalar case, it is straightforward to demonstrate the following:

\begin{lem}\label{lem_mult_group}
the
rescaling factor $\underline{\underline{a}}(n)$
forms a multiplicative group:

\begin{equation}\label{eq_group}
\forall m,n \in N:\quad
\underline{\underline{a}}(m.n)=\underline{\underline{a}}(m)\underline{\underline
{a}}(n)
\end{equation}
\end{lem}

In order to obtain this group property, it suffices to iterate
Eq.\ref{def_levy_vector}
over $m.n$  since it yields:

\begin{equation} \label{eq_iteration}
\underline{\underline{a}}(m.n) \underline{X} +\underline{\gamma}(m.n) =
\underline{\underline{a}}(n). \underline{\underline{a}}(m) \underline{X}
+\underline{\underline{a}}(n). \underline{\gamma}(m) +m\underline{\gamma}(n)
\end{equation}

Lemma \ref{lem_mult_group} yields the following:

\begin{lem}\label{lem_mult_gen}
The multiplicative group $\underline{\underline{a}}(n)$ has a generator
$\underline{\underline{\alpha}}^{-1}$, i.e.:

\begin{equation}\label{eq.generator.alpha}
\exists \underline{\underline{\alpha}} \in L(H,H), \forall n  \in N:
\underline{\underline{a}}(n)= n^{\underline{\underline{\alpha}}^{-1}}
= e^{Ln(n)\underline{\underline{\alpha}}^{-1}}
\end{equation}
\end{lem}

Due its symmetry in $m$ and $n$, Eq.\ref{eq_iteration} also shows that:

\begin{equation}\label{eq.gamma}
\exists \underline{b} \in H, \forall n:
\underline{\gamma}(n)= [\underline{\underline{a}}(n)-n] \underline{b}
\end{equation}

\noindent therefore,  it demonstrates
the following
\begin{lem} \label{reduc_strict}
If $\underline{X}_i$ are stable with generator
$\underline{\underline{\alpha}}^{-1}$
and $1 \notin Spec(\underline{\underline{\alpha}}) $,
there exists $\underline{b}$ (eq.\ref{eq.gamma}), such that
$\underline{X}_i -\underline{b}$
 are strictly stable with the same generator.
\end{lem}

\section{Characteristic functions for stable
vectors}\label{sect_characteristic}

The first  (resp. second) characteristic functions
$Z_{\underline{X}}(\underline{q})$
(resp. $K_{\underline{X}}(\underline{q})$) are defined in the following way:

\begin{equation}\label{eq_vector_charactersitics}
Z_{\underline{X}} (\underline{q}) = e^{K_{\underline{X}}(\underline{q})} =
\int_{H'} {e^{(\underline{q}, \underline{X})}dP_{\underline{X}}}
\end{equation}

\noindent where $P_{\underline{X}}$ is the probability of $\underline{X}$
and the domain $H'$ of the conjugate vector $\underline{q}$ is respectively
$H'= H^+, iH, \overline{H}$ for
Laplace characteristic functions, for Fourier
characteristic functions, and Fourier-Laplace functions\footnote{As for the
scalar case
the Fourier transform is defined for any type of stable vectors, whereas
the Laplace transform
is only defined for extremely asymmetric cases, but are more convenient for
the latter case.}.

With the help of the lem.\ref{reduc_strict}, we need only to consider the
case of
strictly stable vectors. In this case the group property of
\underline{\underline{a}}(n)
(eq.\ref{eq.generator.alpha}) corresponds to:

\begin{theorem}\label{theo_scaling_characteristic}
The second characteristic function $K_X(\underline{q})$ of a strictly stable
stable vector has the following scaling behavior

\begin{equation}\label{eq_scaling_characteristic}
\forall \lambda \in R^+, \forall \underline{q} \in H':
K_{\underline{X}}(\underline{\underline{T}}_{\lambda}.\underline{q}) =
\lambda K_{\underline{X}} (\underline{q})
\end{equation}

\noindent where the dilation $\underline{\underline{T}}_{\lambda}$ has the
generator
$\underline{\underline{\alpha}}^{-1}$.

\end{theorem}

For any positive integer $\lambda$, this is an immediate consequence
of def.\ref{def_levy_vector} and lem.\ref{lem_mult_gen}.
It is readily extended to any inverse of integer $\lambda$, by considering the
intermediate vector $\underline{q}'=
\underline{\underline{T}}_{\lambda^{-1}}.\underline{q}$,
and therefore to any rational $\lambda$. Finally, due to the continuity of
$K_{\underline{X}} (\underline{q})$,
this is true for any positive real $\lambda$. In the classical case, we have:

\begin{equation}\label{classical_norm}
\underline{\underline{\alpha}}= \alpha\underline{\underline{1}}; \quad
\parallel\underline{y}\parallel= \mid\underline{y}\mid^{\alpha}
\end{equation}

\noindent and on the other hand the component $(\underline{u},\underline{X})$
of a stable vector ($\underline{X}$)
along any given direction $\underline{u}$ ($\mid \underline{u} \mid =1$)
is a stable variable. Therefore, the characteristic of the former is obtained
from the one of the latter, with the help of a positive measure
$d\Sigma'(\underline{u})$ of the
directions $\underline{u}$. With the help of the lem.\ref{reduc_strict},
this yields the classical result \cite{Levy, Paulauskas, Nikias}:

\begin{cor}\label{classical_K}
The second characteristic function of a classical (or quasi-scalar) Levy
stable vector corresponds to:

\begin{equation}\label{eq_classical_K}
K_{\underline{X}}(\underline{q})= \int_{\underline{u} \in \partial B'_1}
(\underline{q},\underline{u})^{\alpha}
d\Sigma'(\underline{u})+(\underline{q},\underline{b})
\end{equation}
where $d\Sigma'$ is a positive measure which support $\partial B'_1$ is a
subset of the frontier of the unit ball $\partial B_1$
\footnote{The symmetry of the probability
distribution is related to the one of $d\Sigma$, in the extreme
asymmetric case, the support $\partial B'_1$ of this measure is a subset of
$H^+ \cap \partial B_1$.}.

\end{cor}

\noindent it is straightforward that it is  solution of
eq.\ref{eq_scaling_characteristic}.

Eq.\ref{eq_classical_K} already points out a major difficulty with L\'evy
stable vectors:
a classical L\'evy multivariate distribution is in general not parametric,
contrary to an attempt
to reduce it to a parametric distribution \cite{Press}, which turns out
\cite{Paulauskas} to be
only a very particular sub-case.

One may note that in case of a Fourier characteristic function,
eq.\ref{eq_classical_K} yields the classical expression for $\alpha \in
]0,1[ \cup ]1,2[$:

\begin{equation}\label{eq_classical_K_explicit}
q=iq', q' \in H;~~ K_{\underline{X}}(\underline{q})= i
(\underline{q'},\underline{b})
-\int_{\underline{u} \in \partial B_1}
(\underline{q},\underline{u})^{\alpha} d\Sigma'(\underline{u})
+ i \beta(\underline{q'})
\end{equation}

\noindent where the asymmetry function $\beta(\underline{q})$ is given by:
\begin{equation}
\beta(\underline{q'}) = tan({{\pi \alpha} \over {2}}) \int_{\underline{u}
\in \partial B_1}
(\underline{q'},\underline{u})
\mid(\underline{q'},\underline{u})\mid^{\alpha - 1} d\Sigma'(\underline{u})
\end{equation}

\section {Levy canonical measure and generation of Levy stable vectors}

Let us recall that the Levy 'canonical' measure of a stable random variable
can be best understood as corresponding to the distribution of hyperbolic
jumps
in a Poisson compound process \cite{Levy, Fan}.

This corresponds to substituting a Poisson sum
for the deterministic sum in the definition of stability
(eq.\ref{eq1_levy_variable} for stable variables, respectively
eq.\ref{eq_levy_vector},
for stable  vectors), as well as for the definition of acttractivity
(eq.\ref{eq_attractive_vector}).
The second characteristic function $K_{\underline{X}}(\underline{q})$
is therefore rather easily determined with the help of the  is the Levy
canonical measure
which is the $\sigma -$ probability $F_{\underline{Y}}$
of the jumps $\underline{Y}$.

Indeed, a  compound Poisson process of a random vector field $\underline{Y}$
defined by a measure $d \sigma$, generates
the following  (random) measures $\underline{S}(A)$
for any given borelian  $A$ (of the space on which the process takes place):

\begin{equation}
\underline{S}(A) = \int_{A} \underline{Y}  d \sigma
\end{equation}

\noindent and which have the following characteristic function:

\begin{equation} \label{charactersitic_compound_P}
 K_{\underline{S}(A)} (\underline{q}) = c \sigma(A) \int
[e^{(\underline{q},\underline{y})}
 - \omega((\underline{q},\underline{y}))-1]
dF_ {\underline{y}}
\end{equation}

\noindent where the last term under the integral of the right hand side, is
classical and merely removes
the divergence of the $\sigma -$ probability $F_{\underline{y}}$ in the limit
$\mid\underline{y}\mid \to 0$. Indeed, a  $\sigma -$ probability does need to
be finite, but only to be the limit $\varepsilon \to 0$
of finite positive measures $F_{\underline{y}}^{\varepsilon}$.  The term
$\omega(\underline{q},\underline{y})$
rather corresponds to a generalization of the
'Levy's trick' for removing other divergences
($(\underline{q},\underline{y}) \longrightarrow 0$)
of higher order (i.e. $(\underline{q},\underline{y})$), and whose
appropriate choice will be discussed
below. In order to determine the latter, we have first to determine the
scaling behavior of
the Levy canonical measure $F_{\underline{Y}}$. With no surprise, we  obtain
by conjugation of theorem \ref{theo_scaling_characteristic}:

\begin{theorem}\label{theo_scaling_F}
The Levy canonical measure $F_{\underline{Y}}$ of the jumps $\underline{Y}$
of a stable vector $\underline{X}$ has the following scaling behavior:

\begin{equation}\label{eq_scaling-F}
\forall \lambda \in R^+, \forall \underline{y} \in H:
dF_{\underline{\underline{T}}^{*}_{\lambda}(\underline{y})}= {\lambda}^{-1}
dF_{\underline{y}}
\end{equation}

\noindent where the dilation $\underline{\underline{T}}^{*}_{\lambda}$ is
the conjugate of
$\underline{\underline{T}}_{\lambda}$, i.e. has for generator:
$\underline{\underline{\alpha}}^{*-1}$

\end{theorem}

In order to make a step further, let us consider the following generalized
definition of the 'unitary vector'
$\underline{u}(\underline{y})$ corresponding to a given vector $\underline{y}$:

\begin{definition} \label{def_unitary}
In the framework of a generalized scale (def.\ref{def_GSI}),
the unitary vector $\underline{u}(\underline{y})$ of any non-zero vector
$\underline{y}$
is defined by:
\end{definition}

\begin{equation}
\forall \underline{y} \in H-\{\underline{0}\},
\exists \underline{u}(\underline{y}) \in \partial B_1:
\underline{u}(\underline{y}) =
\underline{\underline{T}}^{-1}_{\parallel\underline{y}\parallel}
(\underline{y})
\end{equation}

In the case of the norm $\mid. \mid$, it corresponds to the usual notion of
unitary vector, i.e.:

\begin{equation}
\underline{u}(\underline{y})= {\underline{y} \over {\mid \underline{y} \mid}}
\end{equation}

\begin{theorem} \label{theo_factorization}

The Levy canonical measure factors into a measure having a density in scale
and  a positive measure $d\Sigma$ on the frontier of unit ball $\partial B_1$:
\end{theorem}

\begin{equation} \label{eq_factorization}
dF_{\underline{y}} = d\Sigma(\underline{u})
{{d  \parallel\underline{y}\parallel} \over
{\parallel\underline{y}\parallel^2}}
\end{equation}

Indeed, due to its scaling property (theorem \ref{theo_scaling_F})
the Levy canonical measure $F_{\underline{Y}}$ should
factor into a scaling measure
of $\parallel\underline{Y}\parallel$ (therefore a power law) and a  measure
invariant for any dilation
$\underline{\underline{T}}^{*}_{\lambda}$. Since every ball $B_{\lambda}$
is obtained by dilation
of the unit ball $B_1$, it suffices to consider a measure on the frontier
of the latter and
to take care of the fact that the Levy canonical measure should be positive.
The classical case (eq.\ref{classical_norm}) does correspond to:

\begin{equation}\label{classical_F}
dF_{\underline{y}}=\alpha {{d\mid y \mid} \over {\mid y \mid}} \mid y
\mid^{-\alpha} d\Sigma(\underline{u})
\end{equation}

\noindent and with the (classical) choice of either $\omega(x)=0, (0<
\alpha \le 1)$ or
$\omega(x)=x, (1< \alpha \le 2)$, this yields eq.\ref{eq_classical_K}
and the proportionality  between the measure
$ d\Sigma'$ (eq.\ref{eq_classical_K}) and  $d\Sigma$ (eq.\ref{classical_F})
is determined with the help of the Euler $\Gamma$ function:

\begin{equation}
d\Sigma' = {\Gamma(3- \alpha) \over \alpha(\alpha-1)}d\Sigma
\end{equation}

As mentioned by Levy, another possibility is to choose
\footnote{One may note that L\'evy used in fact a simplified expression
which is not relevant for
the following}:

\begin{equation} \label{eq_omega}
\omega(x)={x \over {1+x^2}}
\end{equation}
\noindent which is valid for any $\alpha \in ]0,2]$, since $\omega \sim x
(x \longrightarrow 0)$
and $\omega \sim 0 (x \longrightarrow \infty)$.

For the general case, the characteristic function is rather more involved:

\begin{cor} \label{cor_carac_function}
The second characteristic function of a Levy stable vector has the
following expression

\begin{equation}\label{eq_general_K}
K_{\underline{X}}(\underline{q})= \int_{\underline{u} \in \partial B_1}
d\Sigma(\underline{u})
\int_{0}^{\infty} {{d \lambda} \over {\lambda}^2}
 [e^{(\underline{q},\underline{\underline{T}}^{*}_{\lambda} (\underline{u}))}
 - \omega(\underline{q},\underline{\underline{T}}^{*}_{\lambda}
(\underline{u}))-1]
 \end{equation}

\noindent where $\omega$ is defined by eq.\ref{eq_omega},
$d\Sigma$ is a positive measure on the frontier of the unit ball $\partial
B_1$
and the spectrum of $\underline{\underline{\alpha}}$, which is the inverse
of the generator of the
 dilation $\underline{\underline{T}}_{\lambda}$, should satisfy:

\begin{equation} \label{eq_spec_alpha}
  Re(Spec(\underline{\underline{\alpha}})) \subseteq ]0,2]
\end{equation}

\end{cor}

Eq.\ref{eq_spec_alpha} is a rather straightforward extension of the scalar
case constraint
($0 < \alpha ¾ 2$). On the one hand, the lower spectral bound $0$ of
$\underline{\underline{\alpha}}$
allows its inverse to be defined, whereas the upper spectral bound $2$,
which not surprisingly corresponds
to the gaussian limit case, allows the definition of a generalized notion
of scale associated
to the generator $\underline{\underline{\alpha}}^{-1}$
(sect.\ref{sect_GSI} and in particular theo.\ref{cor_existence_gsi})
and it is required in order to ensure the convergence of
the second integration involved in eq.\ref{eq_general_K} for the lower
bound ($\lambda \to 0$).
More precisely the integrand is of order:

\begin{equation}
{{d \lambda} \over {\lambda}^2}
(\underline{q},\underline{\underline{T}}^{*}_{\lambda} (\underline{u}))^2
\end{equation}

\noindent and it should be bounded by below by ${d \lambda} \over
{\lambda}$ to avoid divergences.
We have the following two cases:

\begin{itemize}
\item{\it diagonalizable} $\underline{\underline{\alpha}}$: \\
eq.\ref{eq_spec_alpha} merely corresponds to imposing the adequate
constraint to
each eigenspace, therefore to the full space.

\item{\it otherwise}:\\
one needs to consider the Jordan form of
$\underline{\underline{\alpha}}^{-1}$.
And on each generalized eigenspace corresponding to the eigenvalue
$\alpha^{-1}=a+ib, a,b \in R$
the component (e.g.\cite{Perko}) of
$\underline{y}_{\lambda}=\underline{\underline{T}}_{\lambda}^{*}
(\underline{u}) $
is a linear combination of functions of the form
${\mid Log(\lambda) \mid}^{k}{\lambda}^{a}cos(b\lambda)$ and
${\mid Log(\lambda) \mid}^{k}{\lambda}^{a}sin(b\lambda)$, where $k$ is
bounded above by the codimension of the deficiency of the eigenvalue:
$k \leq c= inf_{n \in N}[codim(Ker((\underline{\underline{T}}_{\lambda}-
\lambda\underline{\underline{1}})^n)]$
Therefore the previous result holds.

\end{itemize}

\section{Conclusion}
We demonstrated that the notion of Levy stable vectors can be broadly
generalized,
the stability index becoming a linear application
$\underline{\underline{\alpha}}$
which needs to respect only one spectral constraint
(eq.\ref{eq_spec_alpha}). We have two main cases:

\begin{itemize}
\item{\it diagonalizable} $\underline{\underline{\alpha}}$: \\
on each of eigenspace, the stable vector have a common stability index
which is
the corresponding eigenvalue. However, there is no need for the eigenvalues
to be equal,
as in fact hypothesized in the classical definition of Levy stable vectors.

\item the generalization is in fact {\it much broader}:\\
indeed complex eigenvalues of $\underline{\underline{\alpha}}$
induce rotation modulations, and deficiency of its eigenvalues introduce
logarithmic modulations. As a
consequence the stability index of a component of a stable vector depends
on its norm.
\end{itemize}

Not only these results are fundamentaly important for multivariate analysis
of random vectors,
but for their stochastic simulations. Indeed, this study was in fact
motivated by the latter, more
precisely by simulations of multifractal fields \cite{fractals}
generated by strongly anisotropic stochastic differential equations. The
latter
naturaly introduce the generalized scale notion used in the present paper.

\section{Acknowledgments}

This paper was supported by the International Association for the Cooperation
with Scientists from the Independent States of the Former Soviet Union
(INTAS-93-1194). We thank H. Gao for a stimulating discussion on Generalized
Scale Invariance.

\end{document}